\documentstyle[aps,prb,preprint]{revtex}
\tighten
\begin{document}
\title{ Three--dimensional Josephson--junction arrays \\ in the quantum
     regime}
\date{\today}
\author{T. K. Kope\'{c}$^{1}$ and Jorge V. Jos\'{e}$^{2}$}
\address{
$^{1}$Institute for Low Temperature and Structure Research,
Polish Academy of Sciences,\\
 POB 1410, 50-950 Wroclaw 2, Poland\\
$^{2}$Physics Department and Center for Interdisciplinary
Research on Complex Systems,\\ Northeastern University,
Boston, MA 02115 USA
}

%\begin{abstract}
%%%%%%%%%%%%%%%%%%%%%%%%%%%%%%%%%%%%%%%%%%%%%%%%%%%%%%%%%%%%%%
\address{\mbox{ }}
\address{\parbox{14cm}{\rm \mbox{ }\mbox{ }
We  study the quantum phase transition  properties of  a 
three--dimensional  periodic array of Josephson junctions
with charging energy that includes both the self and mutual 
junction capacitances. We use the phase fluctuation 
algebra between number and phase operators, given by the 
Euclidean group $E_2$, and we effectively map
the problem  onto a  solvable  quantum generalization of the spherical model.
We obtain a phase diagram as a function of temperature, Josephson coupling
and charging energy. We also analyze  the corresponding  fluctuation 
conductivity and  its universal scaling form in  the vicinity of 
the zero--temperature quantum critical point.  
}}
\address{\mbox{ }}
\address{\parbox{14cm}{\rm \mbox{ }\mbox{ }
}}
\address{\mbox{ }}
%%%%%%%%%%%%%%%%%%%%%%%%%%%%%%%%%%%%%%%%%%%%%%%%%%%%%%%%%%%%%%%
%\end{abstract}
%\pacs{75.50+r,67.40.Db,73.23.Hk}
\maketitle 

\narrowtext
%\twocolumn
\newpage 

There is  significant contemporary interest in quantum critical 
phenomena. Most studies have  been carried out  in two-dimensions
\cite{damle}. There are several systems where theoretical
results have been successfully compared  against experiment
in artificial networks \cite{rojas} and homogeneous 
ultrathin films \cite{films}. There has been some but much less 
work in the three--dimensional (3-D) case,  although there is 
both theoretical and experimental interest in this problem, 
for example, in  quantum magnetic systems and high temperature 
superconductors. There is also preliminary progress 
in fabricating quasi-three dimensional Josephson junction arrays (JJA) with 
ultrasmall junctions, in which quantum fluctuations are essential 
\cite{mooj}.  There is also interest in the classical limit of 
bulk high--$T_c$ superconductors where the scaling critical 
properties are dominated  by thermal fluctuations
\cite{salomon}. Closely related is also the physics that governs the interplay
between local and global superconductivity in granular materials, in
which disorder may also play an important role. In spite of this interest, 
however,  3-D quantum--capacitive JJA have not been investigated in depth yet.
Notably, there appear to be no studies on 3-D JJA close to the $T=0$
quantum--critical (QC) point,  where the physics is dominated
by zero--point quantum fluctuations rather then thermal effects.
When the superconducting islands can sustain at least one Cooper pair
the development of global superconducting  phase coherence
depends on the relative strength of the inter-island Josephson coupling $E_J$,
as compared to the charging energy $E_C=e^2/2C$, where C is the junction
capacitance. In the {\it quantum} regime the phase-charge interplay is a 
direct consequence of the Heisenberg uncertainty  relations between 
the island phase $\phi_j$  and the particle number operator 
$L_j=i\partial/\partial\phi_j$. In this paper we investigate a general  
quantum--capacitive model for 3-D JJA on the simple cubic lattice.
We establish the general phase transition boundary and present results
for the measurable frequency dependent conductivity in the QC regime.
We employ a novel non mean--field approach based on the proper 
quantum phase fluctuation algebra, by mapping  the 3-D JJA model onto an
effectively constrained system -- a solvable quantum spherical model.

We start by defining a cubic Josephson junction array 
with superconducting phases $\phi_i$ at the 3-D lattice sites $i$. 
The corresponding  effective Euclidean action, in the Matsubara 
``imaginary time'' $\tau $ formulation  ($0\le \tau \le 1/k_BT\equiv \beta$,
with $T$ being the temperature) is ${\cal S}[\phi]={\cal S}_C[\phi]
+{\cal S}_J[\phi]$, where
\begin{eqnarray}
{\cal S}_C[\phi]&=&\frac{1}{8e^2}
\sum_{ij}\int_0^\beta d\tau
\left(\frac{\partial\phi_{i} }{\partial\tau}
\right) {C}_{ij}
 \left(\frac{\partial\phi_{j} }{\partial\tau}
\right),
\nonumber\\
{\cal S}_J[\phi]&=&\sum_{\langle ij\rangle}\int_0^\beta d\tau
J_{ij} \left\{1-\cos[\phi_{i}(\tau)-
\phi_{j}(\tau)]\right\}.
\label{action}
\end{eqnarray}
Here ${\cal S}_C[\phi]$ defines the electrostatic energy, with ${C}_{ij}$
being the  geometric capacitance matrix of the array. This matrix is normally 
approximated, both theoretically and in experimental interpretations as:
$C_{ij}=(C_s+zC_m)\delta_{ij} -C_m\sum_{d}\delta_{i,j+ d}$, with the vector
$d$ running over nearest neighbors, with  $C_s$ the self-capacitance and 
$C_m$ the mutual-capacitance between nearest neighbors ( $z$ stands for the 
coordination number). There are more general forms of the full capacitance 
matrix \cite{gibbons}, but in our analysis the mutual  capacitance 
approximation is sufficient. Finally, ${\cal S}_J[\phi]$ gives the 
Josephson energy $E_J$ (with $J_{ij}\equiv E_J$ for $|i-j|=|d|$ and zero otherwise).

Most analytical works on quantum JJA have employed different kinds of 
mean--field--like approximations \cite{simkin,kim,zaikin,simanek}, 
which are not fully reliable to treat  spatial and  temporal 
quantum phase fluctuations. Furthermore, as pointed out recently 
\cite{doniach},   the JJA model (\ref{action}) must encode the 
phase fluctuation algebra given by the Euclidean group $E_2$, that 
involves  the commutation relations between particle $L_j$  and phase 
(ladder) operators $P_j=e^{i\phi_j}$: $[L_i,P_j]=-P_j\delta_{ij}$,
$[L_i,P^\dagger_j]=P_j^\dagger\delta_{ij}$ and $[P_i, P_j]=0$
with the conserved quantity  (invariant of the $E_2$ algebra)
\begin{equation}
P_iP^\dagger_i\equiv P_{xi}^2+P_{yi}^2=1.
\label{e2constr}
\end{equation}
Thus, the proper theoretical treatment of  a quantum JJA
must maintain the constraint (\ref{e2constr}).
To proceed we  write the partition function
$Z=\int\left[\prod_{i}{\cal D}\phi_{i}\right]e^{-S[\phi]]}$ for
the model (\ref{action}) in terms  of its path integral representation
\cite{kleinert},
by introducing the  auxiliary complex fields ${\psi}_{i}(\tau)$,
 which replace the original ladder operators $P_i$.
%%%%%%%%%%%%%%%%%%%%%%%%%%%%%  change %%%%%%%%%%%%%%%%%%%%%%%%%%%%%%%
To proceed, we substitute the ``rigid" $E_2$ constraint
   given in Eq. (\ref{e2constr}), by the weaker spherical closure relation
    $\frac{1}{N}\sum_iP_iP^\dagger_i=1$,  which maintains
    (on average) the original condition of Eq. (\ref{e2constr}).
    This substitution allows us to formulate the problem 
    in terms of an (exactly) soluble
    quantum spherical (QS) model (see Ref. .\onlinecite{vojta}).
%%%%%%%%%%%%%%%%%%%%%%%%%%%% change %%%%%%%%%%%%%%%%%%%%%%%%%%%%%%%%%%%%
By using the Fadeev--Popov method
with the Dirac delta-functional, which  facilitates both  the change
of integration variables and the imposition of the spherical constraint
we obtain:
\begin{eqnarray}
Z&=&\int\left[\prod_{i} {\cal D}{\psi}_{i}
{\cal D}{\psi}^\star_{i}\right]
\delta\left(\sum_{i} |{\psi}_{i}|^2
-N\right)
e^{-{\cal S}_J[\psi]}
\nonumber\\
&\times&\int\left[\prod_{i}{\cal D}\phi_{i}\right]
e^{-{\cal S}_C[\phi]}
\prod_{i} 
\delta\left[{\Re e\psi}_{i}-
{P}_{i}^x(\phi)
\right]\nonumber\\
&\times&
\delta\left[{\Im m}\psi_{i}-
{ P}_{i}^y(\phi)
\right].
\label{changevar}
\end{eqnarray}
The convenient way to enforce the spherical constraint
is to use the functional analog of the $\delta-$function
representation $\delta(x)=\int_{-\infty}^{+\infty}(d\lambda/2\pi)
e^{i\lambda x}$, which introduces the Lagrange multiplier $\lambda(\tau)$
thus adding an additional quadratic term (in the $\psi-$fields)
to the action (\ref{action}). The evaluation of the effective action 
in terms of the $\psi$ fields may be organized using the loop 
expansion method\cite{kopec}. To second order in $\psi_{i}(\tau)$ we obtain
the partition function of the quantum--spherical model $Z\equiv Z_{\rm QS}$:
\begin{eqnarray}
Z_{\rm QS}
=\int\left[\prod_{i} {\cal D}{\psi}_{i}
{\cal D}{\psi}^\star_{i}\right]
\int\left[\frac{ {\cal D}\lambda}{2\pi i}\right]
e^{-{\cal S}_{\rm QS}[
{\mbox{\boldmath$\psi$}},\lambda]},
\label{statsumqsa}
\end{eqnarray}
where
\begin{eqnarray}
&&{\cal S}_{\rm QS}[
{\psi},\lambda]
=\sum_{\langle ij \rangle} \int_0^\beta d\tau d\tau'
\left\{
\left[\left( {J}_{ij}
+\lambda\delta_{ij}\right)
\delta(\tau-\tau')\right.\right.
\nonumber\\
&&+
\left. \left.\Gamma_{ij}(\tau-\tau')\right]
{\psi}^{\star}_{i}(\tau)
{\psi}_{j}(\tau')
-N\delta_{ij}\lambda \delta(\tau-\tau')\right\}.
\label{qsa}
\end{eqnarray}
Here,
$\Gamma_{ij}(\tau-\tau')$
is the two--point phase vertex function related to the
phase--phase cumulant correlation function $W_{kj}(\tau-\tau')$ by
\begin{equation}
\sum_k\int_0^\beta
d\tau''\Gamma_{ik}(\tau-\tau'')W_{kj}(\tau''-\tau')=\delta_{ij}
\delta(\tau-\tau').
\end{equation}
Explicitly, 
\begin{eqnarray}
&&W_{ij}(\tau-\tau')=
\frac{1}{Z_0}\sum_{\{n_{i}\}}\prod_{i}\int_0^{2\pi}
 d\theta(0)\times
 \nonumber\\
&&\int_{\theta(0)}^{\theta(0)+2\pi n_{i}}
{\cal D}\theta_{i}(\tau)
e^{i[\theta_{i}(\tau)-\theta_{j}(\tau')]}
e^{-{\cal S}_C[\theta]} ,
\label{w02}
\end{eqnarray}
where $Z_0$ is the statistical sum of the ``non--interacting" system
described by the action ${\cal S}_C[\theta]$.
Since the values of the phases $\phi_i$ which differ by $2\pi$ are equivalent,
the path integral can be written in terms of the non--compact phase 
variables $\theta_{j}(\tau)$,  defined on the unrestricted 
interval $(-\infty,+\infty)$, 
and by a set of winding numbers $\{n_{j}\}=0,\pm 1,\pm 2,\dots$, which are integers
running from $-\infty$ to $+\infty$ (and physically
reflects the discreteness of the charge \cite{bruder}),
so that $\phi_{j}(\tau)=\theta_{j}(0)
+2\pi i n_{j}\tau/\beta +\theta_{j}(\tau)$.

In the  $N\to\infty$ thermodynamic limit the steepest descents method becomes  exact;
the condition that the integrand in Eq.(\ref{statsumqsa})
has a saddle point $\lambda(\tau)=\lambda_0$ becomes an 
implicit equation for $\lambda_0$:
\begin{equation}
1=\frac{1}{N}\sum_{{\bf k},\omega_\ell}
G({\bf k},\omega_\ell),
\label{gconstr}
\end{equation}
where $G^{-1}({\bf k},\omega_\ell)
=[{\lambda_0-J({\bf k})+2E_C+
{\omega_\ell^2}/{8E_C}}]$
with $\omega_\ell=2\pi\ell/\beta$ $(\ell=0,\pm 1,\pm 2,\dots)$
being the (Bose) Matsubara frequencies and $J({\bf k})$ the Fourier
transform of the Josephson couplings $J_{ij}$, respectively.
As mentioned above, we next proceed by assuming that
$C_{ij}$ has only the nearest--neighbor mutual components.
 For a 3-D simple cubic lattice we obtain for the charging energy
\begin{eqnarray}
E_C&=&\frac{1}{2}e^2[{\bf C}^-1]_{ii}
\nonumber\\
&=&E_{0C}
\frac{ (4-3v_1)^{1/2}(1-v_1)^{-1} }{\pi^2\gamma (C_m/C_s)}
{\bf K}(\kappa_+){\bf K}(\kappa_-),
\end{eqnarray}
where  $E_{0C}=e^2/(2C_s)$ is the charging energy
for the self--capacitive model and ${\bf K}(x)$ stands for  the
complete elliptic integral of the first kind\cite{abramovitz}.
Furthermore,
\begin{eqnarray}
&&\kappa_\pm^2=\frac{1}{2}\pm\frac{1}{4}v_2(4-v_2)^{1/2}-
\frac{1}{4}(2-v_2)(1-v_2)^{1/2}
\nonumber\\
&&v_1=\frac{1}{2\gamma^2}\left[
\gamma^2+3-(\gamma^2-9)^{1/2}(\gamma^2-1)^{1/2},
\right]
\nonumber\\
&&v_2=v_1/(v_1-1),
\end{eqnarray}
where $\gamma=\frac{1}{2}({1+zC_m/C_s})/(C_m/C_s)$.
%
%

%%%%%%%%%%%%%%%%%%%%%%%%%%%%%%%%%%%%%%%%%fig1.eps
%
%
As usual in spherical model calculations the phase boundary (i.e. the
location of the critical points) is determined
by  Eq.(\ref{gconstr}) from the  the upper limit of the eigenvalue spectrum
$\max\{J({\bf k})\}=3E_J$, associated
with  the onset of the  phase transition -- in the spherical model 
the Lagrange multiplier $\lambda_0$ ``sticks'' to that
value at criticality ($\lambda_0=\lambda_0^{\rm crit}=3E_J$)
and stays constant in the whole low temperature phase\cite{spher3}. 
Equivalently, at the critical point
$1/G({\bf k}=0,\omega_\ell=0)=0$. Introducing the density of states
$\rho(E)=\int_{-\pi}^{\pi}{[d^3{\bf k}}/{(2\pi)^3}] 
\delta(E-J({\bf k}))$ we obtain for the critical line
\begin{eqnarray}
1=&&\int_{-\infty}^{+\infty}dE\rho(E)
\sqrt{\frac{2E_C}{3E_J-E}}\times
\nonumber\\
&&\coth
\left[\beta\sqrt{2E_C(3E_J-E)}\right],
\end{eqnarray}
where $\rho(E)\equiv\rho_{s=0}(E)$ and
\begin{eqnarray}
\rho_{s}(E)&=&\frac{1}{\pi^3E_J}
\int^{a_2}_{a_1}dx
\Theta\left(\frac{|E|}{3E_J}-1\right)
\nonumber\\
&\times&
\frac{A_s(x)}{\sqrt{1-x^2}}
{\bf K}\left[\sqrt{1-\left(\frac{E}{2E_J}
+\frac{x}{2}\right)^2}\right]
\label{dens}
\end{eqnarray}
with $a_1=\max(-1,-2-E/E_J)$,
$a_2={\min}(1,2-E/E_J)$;
$\Theta(x)$ is the unit step function
and $A_0(x)=1$.

The current response to an externally applied electromagnetic
field is the conductivity $\sigma$ that is experimentally measurable.
In the context of two--dimensional JJA there are several studies of
$\sigma$  e.g. at the superconductor--Mott--insulator
using   $1/N$ expansion and Monte Carlo analysis (see,
Ref.\onlinecite{cha}),
 the coarse--grained approach\cite{otterlo}
and   an $\epsilon$-expansion\cite{fazio}.

The standard Kubo formula relates the conductivity
to a two--point current--current correlation function. Applying
an external vector potential ${\bf A}$ modifies the  Josephson coupling
by introducing a Peierls phase factor according to:
 $J_{ij}\to J_{ij}\exp(2ei/\hbar c\int_i^j{\bf A}\cdot d{\bf l})$.
The conductivity is obtained as the second derivative of $Z_{QS}$ 
given in Eq.( \ref{statsumqsa}) with
respect to ${\bf A}$.
After performing the derivatives we obtain
(for vanishing magnetic field)  the longitudinal
component of $\sigma(\omega_\nu)\equiv\sigma_{xx}(\omega_\nu,{\bf q}=0)$
as
\begin{eqnarray}
\sigma(\omega_\nu)=&&\frac{2\pi E_J^2}{R_Q\beta\omega_\nu}
\sum_{\omega_\ell}\int_{-\infty}^{+\infty} dE
\bar{\rho}(E)
G(E,\omega_\ell)
\nonumber\\
&&\times\left[
G(E,\omega_\ell)-G(E,\omega_\nu+\omega_\ell),
\right].
\label{conduct2}
\end{eqnarray}
where $R_Q=h/4e^2=6.45 k\Omega$ is the quantum unit of resistance.
We introduced the modified density of states
$\bar{\rho}(E)
=\int_{-\pi}^{\pi}[{d^3{\bf k}}/{(2\pi)^3}]
\sin^2(k_x)\delta(E-J({\bf k}))$, 
with $\bar{\rho}(E)=
\frac{1}{2}\left[\rho_{s=0}(E)
-\rho_{s=2}(E)\right]$, where $\rho_{s}(E)$ is given
in Eq.(\ref{dens}) with $A_2(x)=2x^2-1$.
Evaluating the summation over Matsubara frequencies
and analytically continuing to real frequencies we obtain for the
real part $\sigma'=\sigma'_{\rm sing}+
\sigma'_{\rm reg}$ of the complex dynamic conductivity
(the imaginary part $\sigma''$ can be obtained via the standard
dispersion relation)
\begin{eqnarray}
\sigma'_{\rm sing}(\omega)&=&\delta
\left(\frac{\omega}{\omega_c}\right)
\cdot\frac{\beta\omega_c}{R_Q}
\left(\frac{\pi}{\delta}\right)^2
\int_{-\infty}^{+\infty}dx\eta(x)
\nonumber\\
&\times&\frac{\mbox{cosech}^2\left(\frac{\beta\omega_c}{4}
\sqrt{1-\frac{x-3}{\delta}}\right)}{1-\frac{x-3}{\delta}},
\nonumber\\
\sigma'_{\rm reg}(\omega)&=&\frac{1}{R_Q}
\frac{\pi^2}{4\delta}\left(\frac{\omega_c}{\omega}\right)^2
\eta\left[3+\delta\left(1-\frac{\omega^2}{\omega_c}\right)\right]
\nonumber\\
&\times &
\coth\left(\frac{\beta|\omega|}{4}\right),
\end{eqnarray}
where $\eta(x)\equiv E_J\bar{\rho}(E_Jx)$ and the parameter
 $\delta=\delta_\lambda/E_J$ measures the distance from the critical point
with $\delta_\lambda=\lambda_0-\lambda_0^{\rm crit}$,
and  with $\omega_c=\sqrt{32E_C\delta_\lambda}$
the threshold frequency for particle--hole excitations.

The real part of the conductivity contains two contributions:
first, the Drude weight $\sigma'_{\rm sing}(\omega)$,
is singular since it is proportional to
$\delta(\omega)$ and the regular finite--frequency contribution
to the conductivity, 
 $\sigma'_{\rm reg}(\omega)$, which is due to 
the electromagnetic field induced
transitions to excited states. The singular part in turn is due to
the free charge acceleration. This is so since the
JJA model considered here contains no dissipation mechanism which would 
arise e.g. in the presence of disorder or from a 
coupling of the phase degrees of freedom
to normal electrons (Ohmic damping).
The results of the numerical calculation for the regular 
part of the conductivity
are shown in Fig.\ref{fig3}. 
%
%%%%%%%%%%%%%%%%%%%%%%%%%%%%%%%%%%%%%%%%%fig3.eps

At $T=0$ the singular part vanishes while the regular part can be
evaluated explicitly close to the critical point   with the result
\begin{eqnarray}
\sigma'_{\rm reg}(\omega,\delta\to 0)&=&
\frac{1}{12}\frac{1}{R_Q}\delta^{1/2}
\frac{|\omega|}{\omega_c}\left(1-\frac{\omega^2_c}{\omega^2}
\right)^{3/2}\nonumber\\
&\times&
\Theta\left(\left|\frac{\omega}{\omega_c}\right|-1\right).
\end{eqnarray}
Note, that at the $T=0$ transition,  where the gap in the response
function vanishes ($\delta=0$), there is no universal
dc conductivity as in the two--dimensional case\cite{cha}.
Universality emerges, however,  in a different context. 
We will now present the scaling analysis satisfied  by
$\sigma(\omega)$ in the vicinity of the quantum phase transition
$E_J=E_J^{\rm crit}$, where the temperature 
obeys $0<k_BT<<E_J$. The behavior of the conductivity 
in this regime can be
understood in terms of a universal scaling function that
depends  on a variable which measures the distance of the
superconducting ground state from criticality.
In the quantum critical region we write the spherical constraint
(\ref{gconstr}) in a terms of a low--temperature expansion
\begin{eqnarray}
&&1=\int^{+\infty}_{-\infty}
\eta(x)
\frac{\sqrt{2\alpha}}
{\sqrt{\delta +3 -x}}
\nonumber\\
&&+\frac{2\delta^{1/2}}{\pi^{3/2}\beta E_J}\sum_{\ell=1}^\infty
\frac{1}{\ell}
K_1\left(2\ell\beta E_J\sqrt{\frac{2\delta\alpha}{\pi}}    \right),
\label{lowtemp}
\end{eqnarray}
where $K_1(x)$ is the MacDonald function\cite{abramovitz}
({\it i.e.} second modified Bessel function) and
 $\alpha=E_C/E_J$. Solving Eq.(\ref{lowtemp}) for small
parameter $\delta$ we obtain:
\begin{equation}
\sigma'_{\rm reg}(\omega)=
\frac{1}{12}\frac{1}{R_Q\sqrt{\alpha_c}}\frac{k_BT}{E_J}
F\left(\frac{\omega}{k_BT}\right).
\end{equation}
The temperature $k_BT$ sets also the energy scale
to measure the frequency via the ratio ${\omega}/{k_BT}$.
It is therefore reasonable to introduce the dimensionless scaling variable
$X={\omega}/{k_BT}$ and finally explicitly write the $F$ function as
\begin{equation}
F(X)=\frac{1}{4\sqrt{2}}X\left(1-4\sqrt{\frac{\pi}{3c_2}}\frac{1}{X^2}
\right)^{3/2}\coth(X),
\end{equation}
where $c_2=\int^{+\infty}_{-\infty}dx
{\eta(x)}/
{({3-x})^{3/2}}$.
%%%%%%%%%%%%%%%%%%%%%%%%%%  change %%%%%%%%%%%%%%%%%%%%%%%%%%%%%%%%%%%%%%%
   At criticality the power-law behavior of the model can be deduced
    from Eq.(\ref{gconstr}) by taking the momentum  long wave--length limit.
    By generalizing the analysis to $d$-spatial dimensions 
    we find that the dynamical critical  exponent $z=1$,  and 
    that the correlation length  exponent $\nu=1/(d-1)$, for $d<d_u=3$,
    about the  $T=0$  quantum critical point.  At finite--temperatures,
    close to the quantum phase  transition we obtain $z=1$ and
    $v=1/(d-2)$,  below the upper critical dimension $d_u=4$, respectively
    (the anomalous dimension is in turn  $\eta=0$ for all values of $d$).
%%%%%%%%%%%%%%%%%%%%%%%%%  change %%%%%%%%%%%%%%%%%%%%%%%%%%%%%%%%%%

In conclusion, we have studied a 3-D quantum Josephson
junction array model in the non--perturbative quantum spherical model
approximation. We have explicitly calculated the phase diagram at zero
temperature as well as the conductivity and its scaling properties a low
temperatures about the quantum critical point. There are
several problems left to consider in the future, like the role 
of disorder, dissipation, applied magnetic fields and the impact of anisotropy
(relevant for layered high--temperature superconductors).

This work has been partially supported  by  a NATO Collaborative Research Grant
No. OUTR.CRG 970299, by the Polish Science Committee (KBN)
under grant No. 2P03B--02415 and by NSF grant DMR-9521845.
%%%%%%%%%%%%%%%%%%%%%%%%%%%%%%%%%%%%%%%%%%%%%%%%%%%%%%%%%%%

%%%%%%%%%%%%%%%%%%%%%%%%
\begin{figure}\caption{Charging energy parameter for the 3-D JJA as a 
function of the ratio of mutual $C_m$ and self-capacitance $C_s$.}
\label{fig1}
\end{figure}
\begin{figure}\caption{Phase diagram for a 3-D JJA in the parameter space
defined by temperature $T$ charging energy $E_{0C}$
and the ratio of mutual- to self capacitance
$C_m/C_s$ of a single junction. The system is phase coherent in the region
below the surface.}
\label{fig2}
\end{figure}
\begin{figure}
\caption{Real part of the dynamical conductivity at $T=0$,
for several values of the dimensionless gap parameter $\delta$,
that measures the distance from the critical point:  $\delta=2$,  $\delta=0.5$,
$\delta=0.1$ and $\delta=0.05$ (from the left to right). The arrow
indicates the position of the singular Drude part (vanishing for $T=0$).}
\label{fig3}
\end{figure}

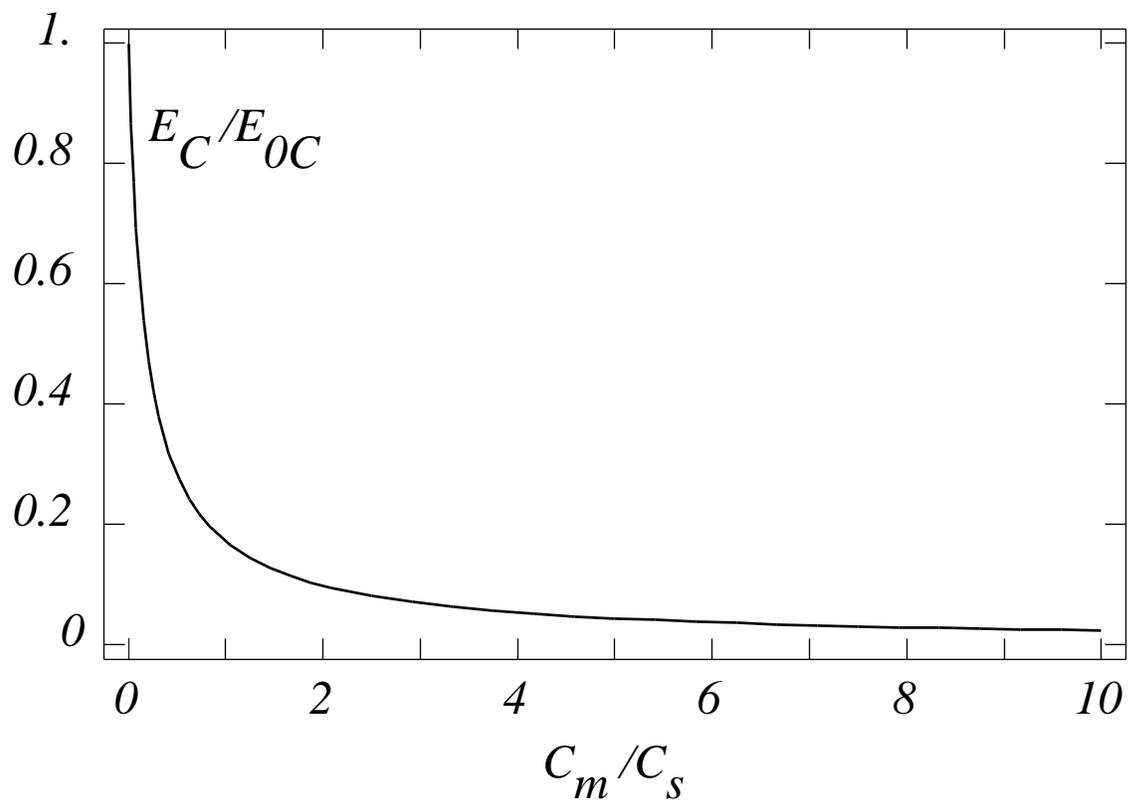
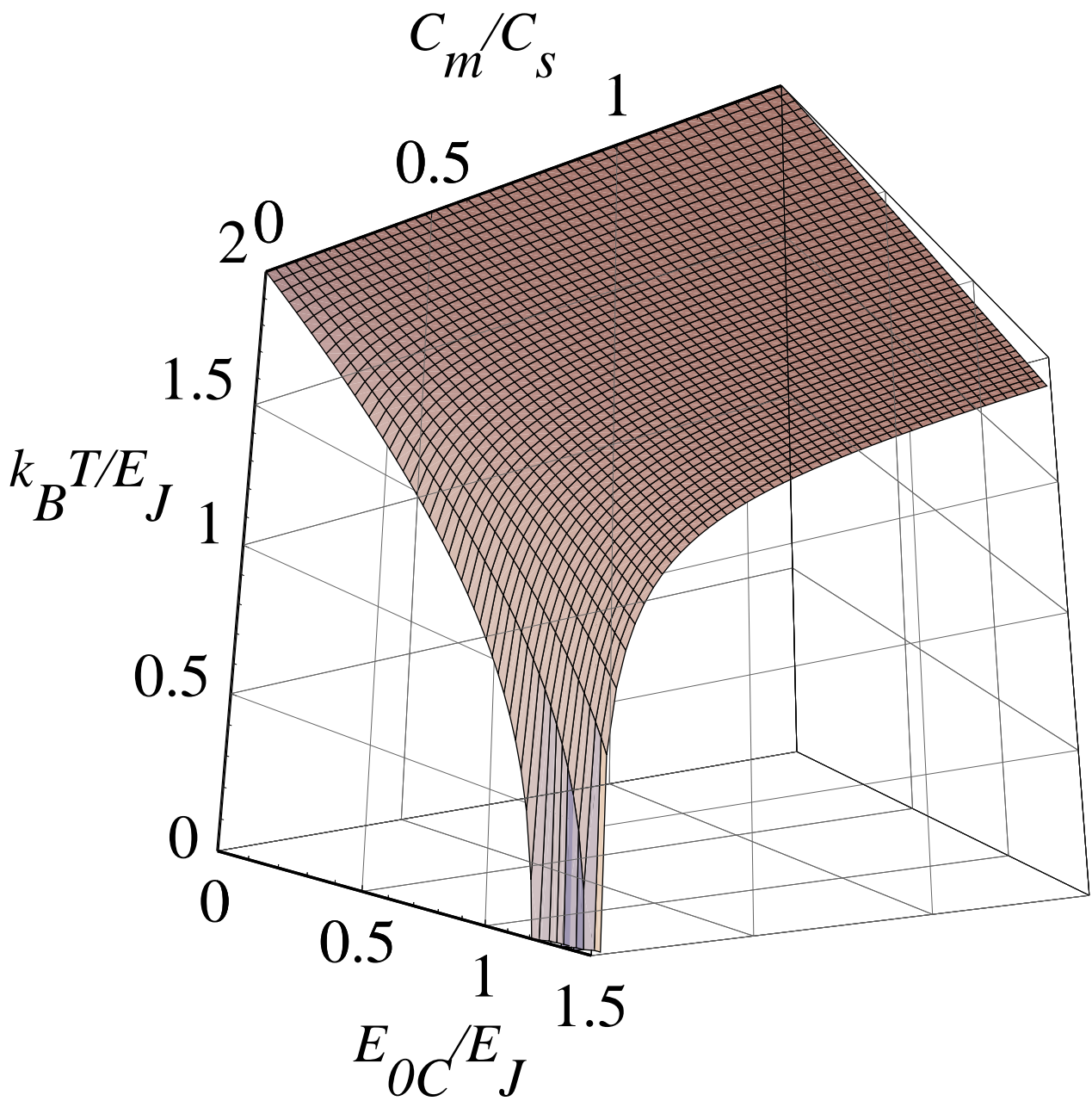
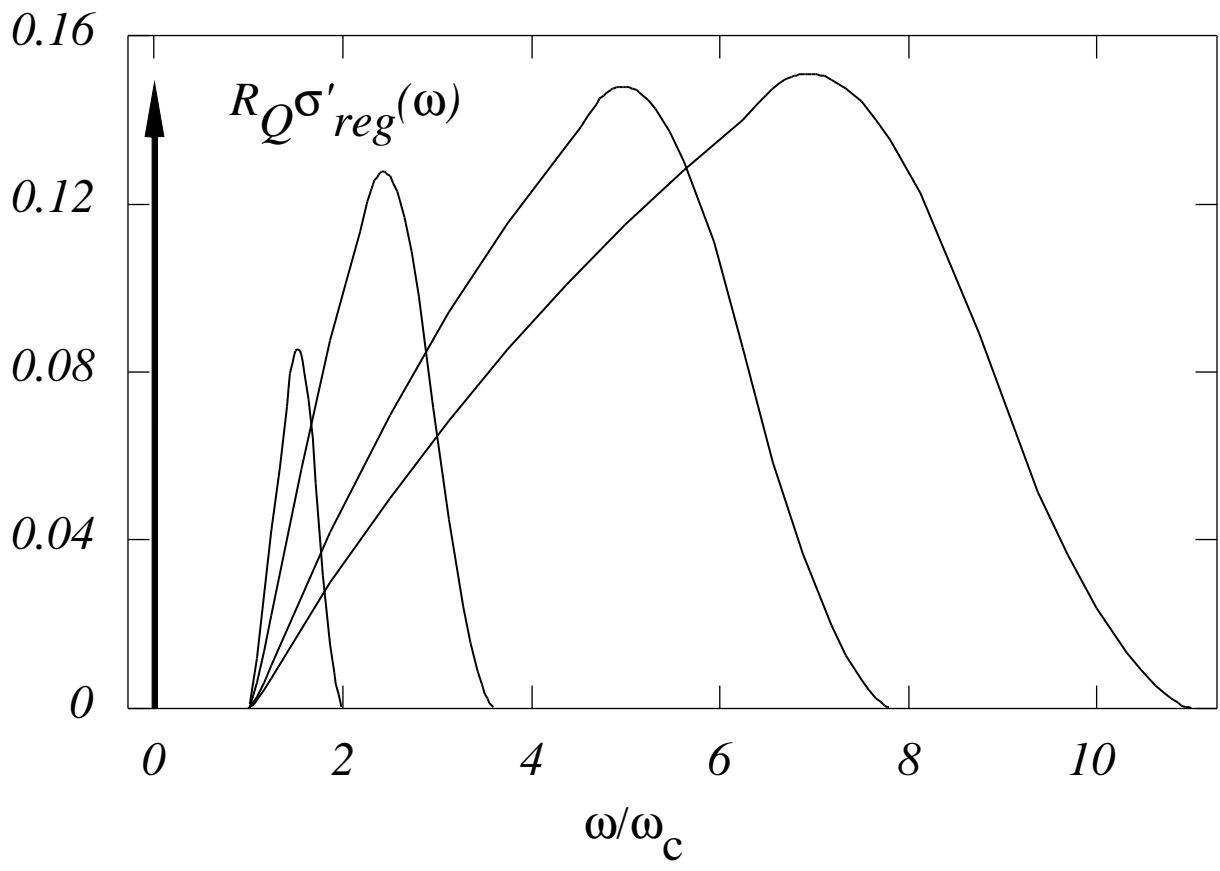
\end{document}